\documentclass[twocolumn,prb,superscriptaddress,floats,aps,amsmath,amssymb]{revtex4-1}
\usepackage[]{graphicx}
\usepackage{color}
\usepackage{bm}
\usepackage{prettyref}
\usepackage{float}
\usepackage{natbib}
\usepackage{tabularx}
\usepackage{booktabs}
\newcolumntype{C}[1]{>{\centering}m{#1}}
\usepackage{array}

\begin{document}
	
	\title{Evidence for $J_{\rm eff} = 0$ ground state and defect-induced spin glass behaviour in the pyrochlore osmate Y$_{2}$Os$_{2}$O$_{7}$}
	\date{\today}
	\author{N. R. Davies}
	\affiliation{Department of Physics, University of Oxford, Clarendon Laboratory, Oxford, OX1 3PU, U.K.}
	\author{C. V. Topping}
	\author{H. Jacobsen}
	\author{A. J. Princep}
	\author{F.~K. K. Kirschner}
	\affiliation{Department of Physics, University of Oxford, Clarendon Laboratory, Oxford, OX1 3PU, U.K.}
	\author{M.~C.~Rahn}
	\thanks{Present affiliation: Los Alamos National Laboratory, Los Alamos, New~Mexico 87545, USA}
	\affiliation{Department of Physics, University of Oxford, Clarendon Laboratory, Oxford, OX1 3PU, U.K.}
	\author{M. Bristow}
	\affiliation{Department of Physics, University of Oxford, Clarendon Laboratory, Oxford, OX1 3PU, U.K.}
	\author{J. G. Vale}
	\affiliation{London Centre for Nanotechnology and Department of Physics and Astronomy, University College London, Gower Street, London WC1E 6BT, United Kingdom}
	\author{I. da Silva}
	\author{P. J. Baker}
	\affiliation{ISIS Facility, STFC Rutherford Appleton Laboratory, Harwell Campus, Didcot OX11 0QX, U.K.}
	\author{Ch. J. Sahle}
	\affiliation{European Synchrotron Radiation Facility, 71 Avenue des Martyrs, 38000 Grenoble, France}
	\author{Y.-F.~Guo}
	\affiliation{School of Physical Science and Technology, ShanghaiTech University, 319 Yueyang Road, Shanghai 200031, China}
	\author{D.-Y. Yan}
	\author{Y.-G. Shi}
	\affiliation{Beijing National Laboratory for Condensed Matter Physics \& Institute of Physics, Chinese Academy of Science, Beijing 100190, China}
	\author{S. J. Blundell}
	\affiliation{Department of Physics, University of Oxford, Clarendon Laboratory, Oxford, OX1 3PU, U.K.}
	\author{D. F. McMorrow}
	\affiliation{London Centre for Nanotechnology and Department of Physics and Astronomy, University College London, Gower Street, London WC1E 6BT, United Kingdom}
	\author{A. T. Boothroyd}
	\email{a.boothroyd@physics.ox.ac.uk}
	\affiliation{Department of Physics, University of Oxford, Clarendon Laboratory, Oxford, OX1 3PU, U.K.}

\begin{abstract}
	
	We present AC and DC magnetometry, heat capacity, muon spin relaxation ($\mu$SR) and resonant inelastic X-ray scattering (RIXS) studies of the pyrochlore osmate Y$_{2}$Os$_{2}$O$_{7}$. We observe a non-zero effective moment governed by $\sqrt{f}\mu_{\rm{eff}} = 0.417(1)\,\mu_{\rm{B}}$ where $f$ is the fraction of Os sites which exhibit a spin, and spin freezing at temperature $T_{\rm f} \simeq 5$\,K, consistent with previous results. The field dependence of magnetisation data shows that the paramagnetic moment is most likely due to large moments $\mu_{\rm eff} \simeq 3\,\mu_{\rm B}$ on only a small fraction $f \simeq 0.02$ of Os sites. Comparison of single-ion energy level calculations with the RIXS data yields a non-magnetic $J_{\rm eff} = 0$ ground state on the Os$^{4+}$ sites. The spin-orbit interaction, Hund's coupling and trigonal distortion of OsO$_{6}$ octahedra are all important in modelling the experimentally observed spectra. We are able to rule out impurity effects, leaving disorder-related effects such as oxygen non-stoichiometry or site interchange between Os and Y ions as the most plausible explanation for the magnetic response in this material.

\end{abstract}

\maketitle

\section{Introduction}

\begin{figure*}
	\centering
	\includegraphics[width=0.9\textwidth]{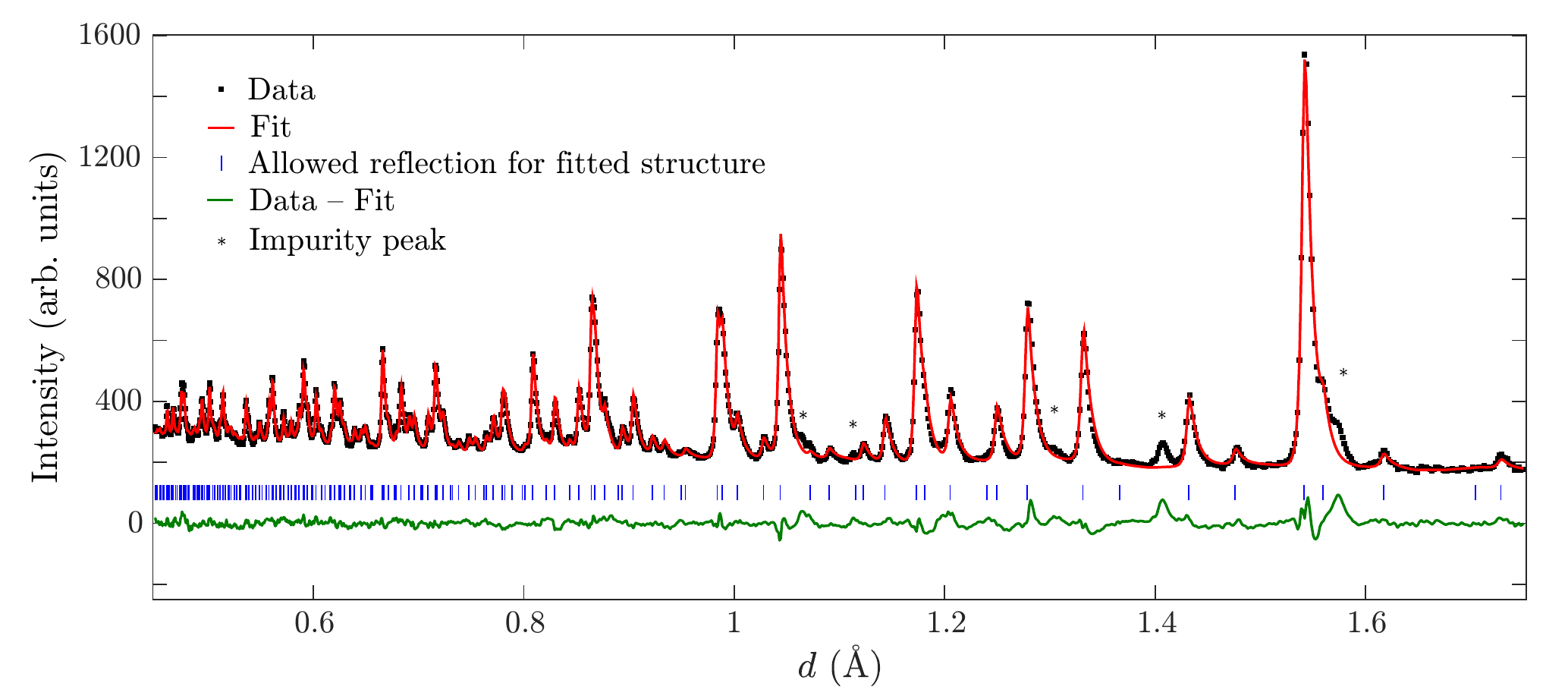}
	\caption{Powder neutron diffraction from Y$_{2}$Os$_{2}$O$_{7}$ measured at a temperature of 200\,K as a function of $d = 2\pi/Q$, where $Q$ is the momentum transfer. Black squares are experimental data and the red line is the fitted profile after structural refinement (corresponding to the 200\,K parameters in Table~\ref{Refinement_Params_Table}). The green line beneath the ticks is a difference plot between data and fit. Blue tick marks show the expected locations of peaks due to the main Y$_{2}$Os$_{2}$O$_{7}$ phase while black asterisks indicate impurity peaks. When regions in which impurity peaks are present are excluded, the Bragg R-factor for this fit is 6.54.}
	\label{GEM_Fig}
\end{figure*}

In the single-ion picture, octahedrally coordinated transition-metal ions with a $d^{4}$ electronic configuration, such as Os$^{4+}$ and Ir$^{5+}$, are expected to have a non-magnetic singlet ground state. For strong spin-orbit coupling, the $t_{\rm 2g}$ levels are split into a fully-filled $j_{\rm{eff}}=3/2$ quadruplet and an empty $j_{\rm{eff}}=1/2$ doublet yielding an overall $J_{\rm{eff}} = 0$, while for strong Hund's coupling each site is in a $S=1$, $L_{\rm{eff}}=1$ state with $L_{\rm{eff}}$ and $S$ coupled by the spin--orbit interaction in a $J_{\rm{eff}} = 0$ state. Since the $d^{4}$ ion is non-magnetic in both of these limits, the singlet ground state is expected to be robust.

Such materials have been studied since the 1960s \cite{Earnshaw1961}, and although many are non-magnetic there are cases in which a magnetic moment and possibly magnetic ordering is nevertheless observed experimentally \cite{Ramos1995, Wang2014, Cao2014, Corredor2017, Dey2016, Hammerath2017, Zhao2016, Ranjbar2015, Terzic2017}, with several different novel mechanisms being proposed to explain this \cite{Khaliullin2013, Metei2015, Bhowal2015, Chen2017}. Notable examples which have been studied recently include the double perovskite iridates $A_{2}$YIrO$_{6}$ ($A = $ Sr, Ba) \cite{Cao2014, Dey2016, Corredor2017} and the pyrochlore osmates $R_{2}$Os$_{2}$O$_{7}$ ($R = $ rare earth) \cite{Zhao2016}.
For the iridates, some theories proposed that a novel excitonic mechanism related to the interplay of spin--orbit coupling and superexchange was behind the magnetic state \cite{Khaliullin2013, Metei2015, Bhowal2015}. It has been pointed out, however, that the superexchange interaction is probably not strong enough in the $A_2$YIrO$_6$ family to induce excitonic magnetism since the IrO$_{6}$ octahedra are isolated from one another \cite{Pajskr2016}. Instead, the observed moment has been ascribed to extrinsic effects, such as paramagnetic impurities \cite{Dey2016, Hammerath2017, Fuchs2018} and antisite disorder \cite{Chen2017, Fuchs2018}.

The pyrochlore osmates provide more promising candidates for excitonic magnetism since the OsO$_{6}$ octahedra form a corner-sharing network, meaning the superexchange is expected to be much larger. Less work has been done on this family of materials, with one experimental study on the $R = $ Y and Ho pyrochlores observing non-zero moments in both cases \cite{Zhao2016}.

In this work we add to previous studies on candidates for excitonic magnetism by reporting measurements on Y$_{2}$Os$_{2}$O$_{7}$ made with a variety of techniques including AC and DC magnetic measurements, muon spin relaxation ($\mu$SR) and resonant inelastic x-ray scattering (RIXS). The interpretation of the data is informed by single-ion electronic structure calculations. We find that the observed paramagnetic moments undergo a bulk spin freezing at low temperatures similar to that found to occur in canonical spin glasses. Single-ion calculations based on the actual distorted crystal structure yield a $J_{\rm eff} = 0$ state in the intermediate coupling regime and indicate that spin--orbit coupling, Hund's coupling and the trigonal crystallographic distortion are all important in modelling the experimental spectra. Having ruled out impurity effects, the field-dependence of magnetisation measurements shows that the magnetism must be due to large moments $\mu_{\rm eff} \simeq 3\,\mu_{\rm B}$ on only a small fraction $f \simeq 0.02$ of Os sites, allowing us to conclude that the observed paramagnetic moment is likely related to crystalline disorder, such as oxygen non-stoichiometry or site mixing between Os and Y ions.

\section{Experimental details}

\begin{table*}
	\caption{Refined structural parameters for Y$_{2}$Os$_{2}$O$_{7}$ in the space group $Fd\bar{3}m$. The numbers in parentheses are uncertainties on the refinement procedure. There is only one free fractional coordinate in this structure for the O on the $48f$ site ($x_{48f}$), corresponding to trigonal distortion of oxgygen octahedra around the Os site. Zero trigonal distortion corresponds to $x_{48f} = 5/16 = 0.3125$. \\}
	\label{Refinement_Params_Table}
	\begin{tabular}{C{2.5cm}|C{1.4cm}C{1.4cm}C{3cm}C{2cm}C{1.9cm}C{2.3cm}C{2.1cm}}
		\hline
		\hline
		\\[-0.23cm]
		Temperature (K) & $a$ (\AA)& $x_{48f}$& O$_{48f}$ occupancy (\%) & $B_{\rm iso}$(Os) (\AA$^{2}$) & $B_{\rm iso}$(Y) (\AA$^{2}$) & $B_{\rm iso}$(O$_{48f}$) (\AA$^{2}$) & $B_{\rm iso}$(O$_{8b}$) (\AA$^{2}$) \tabularnewline[3pt]
		\hline\\[-0.25cm]
		200 &10.225(1) &0.3352(2) &98.7(5)& 0.46(2) & 0.84(4) & 0.91(3) & 0.78(7) \tabularnewline[2pt]
		100 & 10.222(1)& 0.3354(2)& 98.5(8) & 0.42(3) & 0.71(5) & 0.83(5) & 0.70(9) \tabularnewline[2pt]
		2 & 10.220(1)& 0.3355(2)& 98.5(8) & 0.41(3) & 0.68(5) & 0.81(5) & 0.67(9) \tabularnewline[2pt]
		\hline
		\hline
	\end{tabular}
\end{table*}

A 5.2\,g polycrystalline sample of nominal composition Y$_{2}$Os$_{2}$O$_{7}$ was synthesised through a conventional solid-state reaction. A stoichiometric mixture of Y$_{2}$O$_{3}$ and OsO$_{2}$ was ground and sealed in an evacuated quartz tube, then the tube was heated up to 773\,K and left at this temperature for 24 hours. The product was reground and pressed into a pellet and sealed in a new quartz tube under vacuum. The quartz tube was heated slowly up to 1173\,K and kept at this temperature for 2 days. The target phase was thus obtained after shutting off the furnace.

Elastic neutron scattering measurements performed on the GEM beamline at the ISIS facility \cite{Williams1998} allowed for a full structural refinement as shown in Fig.~\ref{GEM_Fig} and Table~\ref{Refinement_Params_Table}, yielding lattice parameter $a$ = 10.225(1)\,\AA. Superficially, the refinement indicates an oxygen deficiency of approximately 1.5\,\%, but this result is not reliable as there is an uncertainty of around 2\,\% in the neutron scattering length of Os \cite{Sears1992}. The only fractional coordinate in this structure which is not constrained by symmetry is the $48f$ oxygen site $x$-coordinate which we find to be $x_{48f} = $ 0.3352(2) at 200\,K. For reference, zero trigonal distortion of the OsO$_6$ octahedra corresponds to $x_{48f} = 5/16 = 0.3125$, with $x_{48f} > 5/16$ indicating trigonal compression of the octahedra in this case. The structural parameters are similar but not identical to those reported in Ref.~\onlinecite{Zhao2016}, with the variation possibly related to different levels of microscopic disorder resulting from the different sample synthesis routes. A small number of low-intensity peaks from an unidentified impurity phase can also be seen. These peaks could not be indexed by Y, Os, any known oxide of Y or Os, or any material expected to be close to the beam path. Based on the intensity of the strongest peaks in the neutron scattering spectrum we estimate that the impurity is on the level of $\lesssim6\%$ and note that the $\mu$SR, RIXS and specific heat measurements presented in this work, being bulk probes, are expected to be relatively insensitive to this level of impurity.  The potential effects of the impurity on magnetisation will be discussed in Section~\ref{DC_Magnetisation}.

DC magnetisation measurements up to 16\,T and specific heat measurements were performed on a Quantum Design Physical Property Measurement System (PPMS), and AC and DC magnetisation measurements up to 7\,T were performed on a Quantum Design Magnetic Property Measurement System (MPMS). Muon-spin relaxation ($\mu$SR) measurements were performed in a $^{4}$He cryostat (3.8--225\,K) and a dilution refrigerator (92\,mK--3.8\,K) on the MuSR beamline at the ISIS Pulsed Muon Facility \cite{King2013} on part of the powder sample packed in a 25-$\mu$m silver foil packet mounted on a silver backing plate. An additional measurement on the same sample was performed in a $^4$He cryostat on the GPS spectrometer at the Paul Scherrer Institute (PSI) to check the low decay time spectrum at 1.5\,K. Resonant Inelastic X-ray Scattering (RIXS) data were taken on a pressed pellet of the sample at the beamline ID20 of the European Synchrotron Radiation~Facility \cite{MorettiSala2018}. 

\section{Results}

\subsection{DC Magnetisation}
\label{DC_Magnetisation}

\begin{figure*}
	\centering
	\includegraphics[width=0.98\textwidth]{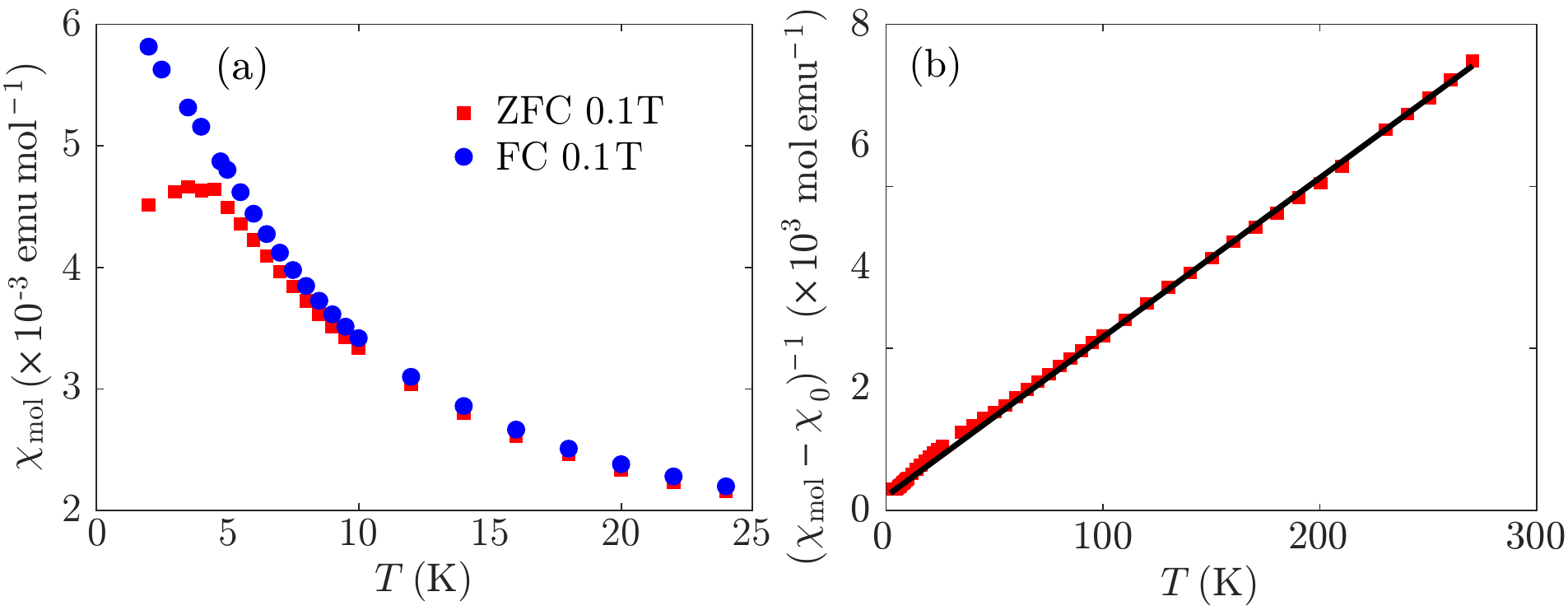}
	\caption{(a) DC magnetic susceptibility of polycrystalline Y$_{2}$Os$_{2}$O$_{7}$. Data were taken using a 0.1\,T measurement field, after cooling from room temperature in the measurement field (FC, blue circles) and in zero field (ZFC, red squares). (b) $(\chi_{\rm{mol}}-\chi_{0})^{-1}$ as a function of temperature, where $\chi_{0}$ is the background susceptibility obtained in a Curie-Weiss fit of the form shown in Eq.~(\ref{CW_Eqn}). The black line is the result of this Curie-Weiss fit, showing good agreement with the data down to around 60\,K. The resulting fit parameters are $\sqrt{f}\mu_{\rm{eff}} = 0.417(1)\,\mu_{\rm{B}}$, $\theta = -2.1(4)$\,K and $\chi_{0} = 8.96(1)\times10^{-4}$\,emu mol$^{-1}$.}
	\label{DC_Fig}
\end{figure*}	

DC magnetic susceptibility data [Fig.~\ref{DC_Fig}] show Curie-Weiss-like behaviour over a large temperature range with departures below about 60\,K and a significant splitting between field-cooled and zero-field-cooled curves below about 5\,K. For temperatures above 70\,K the data fits well to the form
\begin{equation}
\chi = \chi_{0} + \frac{N_{\rm{A}}f\mu_{\rm{eff}}^{2}}{30k_{\rm{B}}(T-\theta)}
\label{CW_Eqn}
\end{equation}
where $\chi$ is the DC susceptibility expressed in CGS units (emu\,mol$^{-1}$), $f$ is the fraction of Os sites which exhibit a magnetic moment and $\mu_{\rm eff}$ is the effective moment due to each of these Os sites [Fig.~\ref{DC_Fig}(b)]. This yields $\sqrt{f}\mu_{\rm{eff}} = 0.417(1)\,\mu_{\rm{B}}$, a Curie-Weiss temperature of $\theta = -2.1(4)$\,K, and a temperature-independent background susceptibility $\chi_{0} = 8.96(1)\times10^{-4}$\,emu\,mol$^{-1}$. The formula unit for all molar quantities here and throughout this work is Y$_{2}$Os$_{2}$O$_{7}$ unless otherwise stated. The origin of the significant temperature-independent component of the susceptibility will be discussed in more detail in Section~\ref{RIXS_Single_Ion} in light of our RIXS results.

The observed DC magnetic susceptibility appears qualitatively very similar to that reported in Ref.~\onlinecite{Zhao2016}, differing only by a constant factor close to 1. The sample measured in that work contained 7\% Y$_{2}$O$_{3}$ and 10\% OsO$_{2}$ impurities and the authors do not mention any unknown impurity similar to the one seen here. The close similarity between the two samples from different sources and containing different secondary phases (albeit at low levels) indicates that the dominant features in the measured magnetisation are from the Y$_2$Os$_2$O$_7$ phase. We will provide further evidence that the impurity does not affect the measured magnetisation in Section~\ref{Analysis} via in-depth analysis of our $\mu$SR results.

\begin{figure*}
	\centering
	\includegraphics[width=0.98\textwidth]{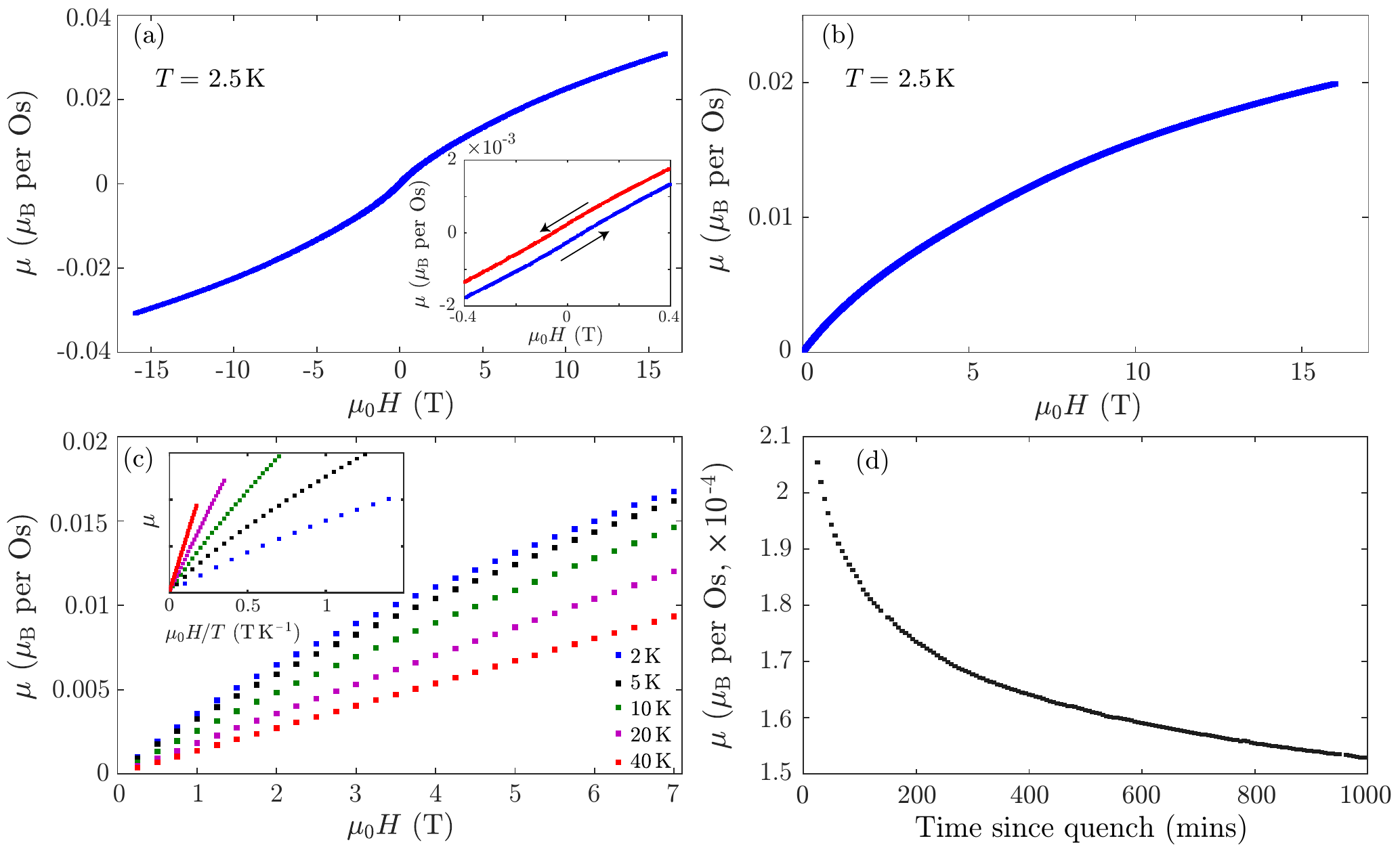}
	\caption{(a) Magnetization $\mu$ of Y$_{2}$Os$_{2}$O$_{7}$ at 2.5\,K as a function of applied field $\mu_{0}H$. A small amount of hysteresis is seen between field up and down sweeps (inset). (b) The positive-field part of the dataset in (a) after subtraction of a linear van-Vleck contribution $\chi_0 = 8.96(1) \times 10^{-4}$\,emu\,mol$^{-1}$, consistent with the results of our Curie-Weiss fitting. (c) The temperature dependence of the magnetization between 2\,K and 40\,K. The inset shows the same data as a function of $\mu_0 H/T$ to demonstrate the lack of $H/T$ scaling. (d) Remanent magnetization of Y$_{2}$Os$_{2}$O$_{7}$ at 2\,K after quenching from a field of 7\,T. In order to obtain this data, the superconducting magnet was ramped down from 7\,T as quickly as possible over the course of around 6 minutes with the starting time $t = 0$ being when the field began to ramp down. The magnet was then heated up above its superconducting transition temperature and the first data point taken once the magnet reached its normal state.}
	\label{MH_Fig}
\end{figure*}

Magnetisation data taken up to 16\,T at 2.5\,K [Fig.~\ref{MH_Fig}(a)] shows a small hysteresis (inset of figure). Due to the significant van-Vleck susceptibility of this material, the moment $\mu$ is expected to be linear in $H$ at high enough fields once the Curie-Weiss-like moments have reached saturation. We therefore subtract a van-Vleck contribution corresponding to $\chi_{0} = 8.96(1)\times10^{-4}$\,emu\,mol$^{-1}$ consistent with the Curie--Weiss fit [Fig.~\ref{MH_Fig}(b)]. The moment does not saturate up to the maximum field of 16\,T, but it does appear to be approaching saturation. We find that the observed rate of approach to saturation makes it unlikely that the saturated paramagnetic moment $\mu_{\rm sat}$ exceeds 0.04\,$\mu_{\rm B}$ per Os site averaged over the whole sample.

The temperature dependence of the magnetisation curves [Fig.~\ref{MH_Fig}(c)] shows that the moment does not exhibit normal paramagnetic behaviour at low temperature. Instead, it is close to temperature-independent at the lowest temperatures (2\,K--5\,K) with ideal paramagnetic behaviour according to the Brillouin function (i.e. $\mu$ being a function of $B/T$ only) not recovered up to 40\,K.

The sample also shows a small but observable remanent magnetisation and time-dependent relaxation when quenched from 7\,T [Fig.~\ref{MH_Fig}(d)] at 2\,K, with the moment not decaying fully even after many hours. The curve does not fit to a single exponential, consistent with a spread of decay times.

\subsection{AC Magnetisation}

\begin{figure*}
	\centering
	\includegraphics[width=0.98\textwidth]{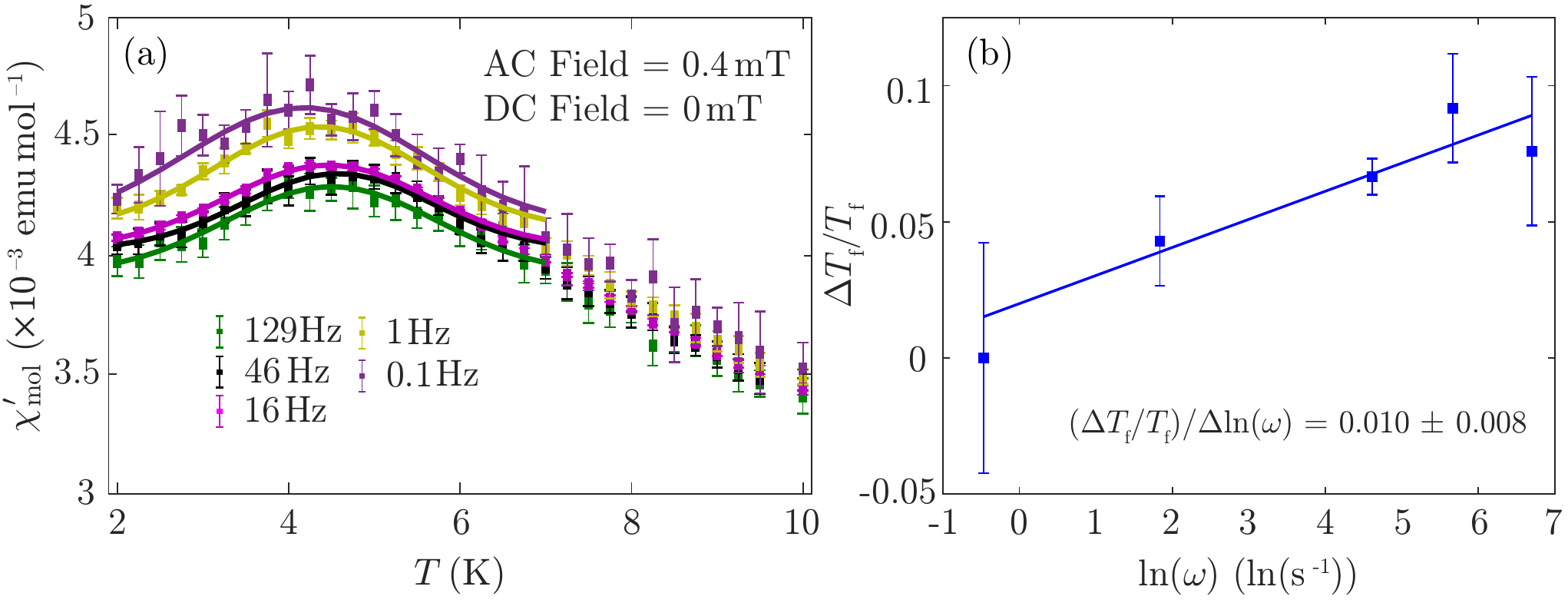}
	\caption{(a) AC magnetic susceptibility of Y$_{2}$Os$_{2}$O$_{7}$ powder. Data were taken using a 0.4\,mT AC measurement field and zero DC field after cooling from room temperature in zero field. $\chi'_{\rm{mol}}$ is the real part of the AC molar susceptibility in CGS units. Solid lines are fits to a Gaussian peak plus a constant background in the region around 4\,K to find the peak position for plotting in (b) and to emphasise its shift to lower temperature with decreasing frequency. (b) Fractional change in the peak position in the real part of the AC magnetic susceptibility $\Delta T_{\rm f}/T_{\rm f,0} = (T_{\rm f}(\omega)-T_{\rm f}(\omega \to 0))/T_{\rm f}(\omega \to 0)$ as a function of $\ln(\omega)$, where $T_{\rm f}(\omega)$ is taken from the fitted peaks in (a). The blue line is the best linear fit yielding a gradient $\Delta T_{\rm f}/T_{\rm f}/\Delta\ln(\omega) = 0.010\pm0.008 $, comparable to that seen in a typical spin glass \cite{Gatteschi2006}.}
	\label{AC_Fig}
\end{figure*}

\begin{figure*}
	\centering
	\includegraphics[width=0.98\textwidth]{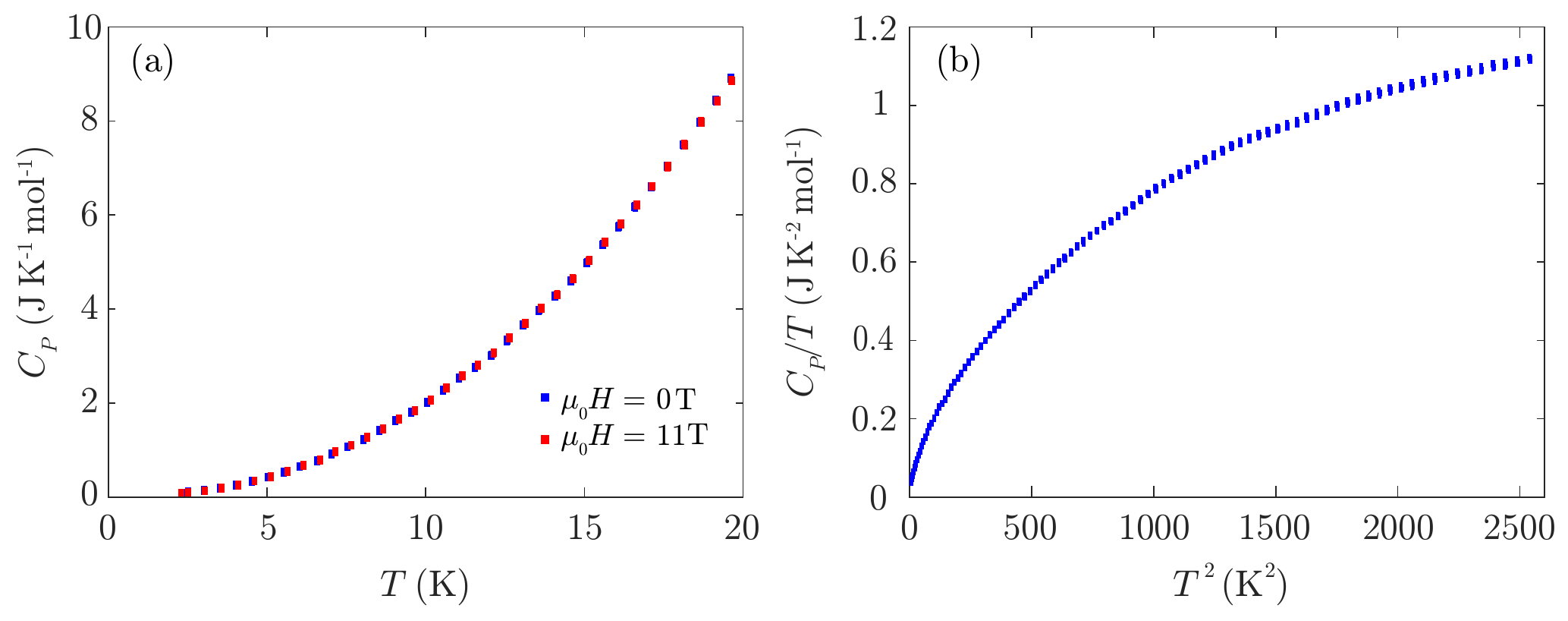}
	\caption{(a) Molar specific heat of Y$_{2}$Os$_{2}$O$_{7}$ as a function of temperature in zero and high (11\,T) magnetic field. (b) $C/T$ as a function of $T^{2}$ in zero magnetic field.}
	\label{Sp_Heat_Fig}
\end{figure*}

In the real part of the AC susceptibility [Fig.~\ref{AC_Fig}(a)] a clear peak is seen close to the proposed spin freezing temperature $T_{\rm f} \simeq 5$\,K (Ref.~\onlinecite{Zhao2016}), with the peak moving to lower temperatures and having a higher maximum $\chi'$ at lower frequencies. This shift in the peak position with frequency shows that there are slow magnetic dynamics in the 0.1--1000\,Hz range. More quantitatively, we find that the peak shift is consistent with the relation
\begin{equation}
\begin{aligned}
\Delta T_{\rm f}/T_{\rm f,0} &= (T_{\rm f}(\omega)-T_{\rm f}(\omega \to 0))/T_{\rm f}(\omega \to 0)\\
&= F\Delta(\ln\omega)
\end{aligned}
\end{equation}
with $F = 0.010(8)$  [Fig.~\ref{AC_Fig}(b)], which is within the range $F = 0.001-0.08$ found for typical spin glasses \cite{Gatteschi2006, Balanda2013}.

We found that the imaginary part of the AC susceptibility was smaller than the instrumental resolution of the magnetometer used for our measurements at all frequencies and temperatures measured, i.e. $\chi'' \lesssim 10^{-4}$\,emu\,mol$^{-1}$. This weak $\chi''$ is consistent with the behaviour of known spin glasses and indicates a wide spread of relaxation times \cite{Balanda2013}.

The hysteretic and frequency-dependent effects described here, including the splitting between field-cooled and zero-field-cooled DC magnetization, AC magnetization and remanence are all characteristic features of canonical spin glasses and other spin-glass-like pyrochlore systems such as Y$_{2}$Mo$_{2}$O$_{7}$ \cite{Raju1992, Dunsiger1996}.

\subsection{Heat Capacity}

The zero-field specific heat of a pressed pellet made from the above sample is smooth at all temperatures down to 2\,K and shows no obvious signature of the spin glass transition or any other magnetic behaviour [Fig.~\ref{Sp_Heat_Fig}(a)]. A plot of $C/T$ as a function of $T^{2}$ [Fig.~\ref{Sp_Heat_Fig}(b)] shows that the data do not fit a simple Debye model ($C/T = \gamma + \alpha T^{2}$) over any measured temperature range. 

Remarkably, on applying a large (11\,T) magnetic field we find no observable change in the specific heat of Y$_{2}$Os$_{2}$O$_{7}$ at any temperature as shown in Fig.~\ref{Sp_Heat_Fig}(a). Since such a large magnetic field can reasonably be expected to significantly affect the magnetic state -- and hence the magnetic component of the heat capacity -- it is very likely that the specific heat measured experimentally is almost entirely due to phonons, and any magnetic contribution is unresolvably small at all temperatures. 

Although surprising, we find that the lack of an observed magnetic specific heat signal is consistent with the results of the other measurements presented here. The effective moment per Os ($\sqrt{f}\mu_{\rm eff}$) is quite small and the spin glass state likely has a large amount of residual disorder, so the entropy change associated with the spin-freezing transition may be quite low. Additionally, the release of entropy for typical spin glasses has been observed to be spread over a large temperature range up to around $5\,T_{\rm f}$ \cite{Binder1986}, resulting in a very small contribution to the specific heat at any given temperature \cite{Wenger1976}.

Our measurements of Y$_{2}$Os$_{2}$O$_{7}$ are consistent with the data presented by Zhou \textit{et al.} in Ref.~\onlinecite{Zhao2016}. However, our conclusion that there is no observable magnetic contribution to the specific heat differs. We therefore performed further heat capacity measurements on a pellet of Y$_2$Ti$_2$O$_7$ as a non-magnetic reference sample. For Y$_2$Ti$_2$O$_7$, we obtained virtually identical data to Zhou \textit{et al.} up to 30\,K. However we find that, after applying the same scaling, the zero-field specific heats of Y$_2$Ti$_2$O$_7$ and Y$_2$Os$_2$O$_7$ are not the same above 30\,K, where any magnetic signal due to spin glass behaviour should be small. The discrepancy between this finding and the conclusions of Ref.~\onlinecite{Zhao2016} indicates that Y$_{2}$Ti$_{2}$O$_{7}$ is not a sufficiently accurate non-magnetic background sample to isolate the small magnetic contribution to the heat capacity.

\subsection{Muon Spin Relaxation ($\mu$SR)} \label{MuSRSection}

\begin{figure}
	\centering
	\includegraphics[width=0.48\textwidth]{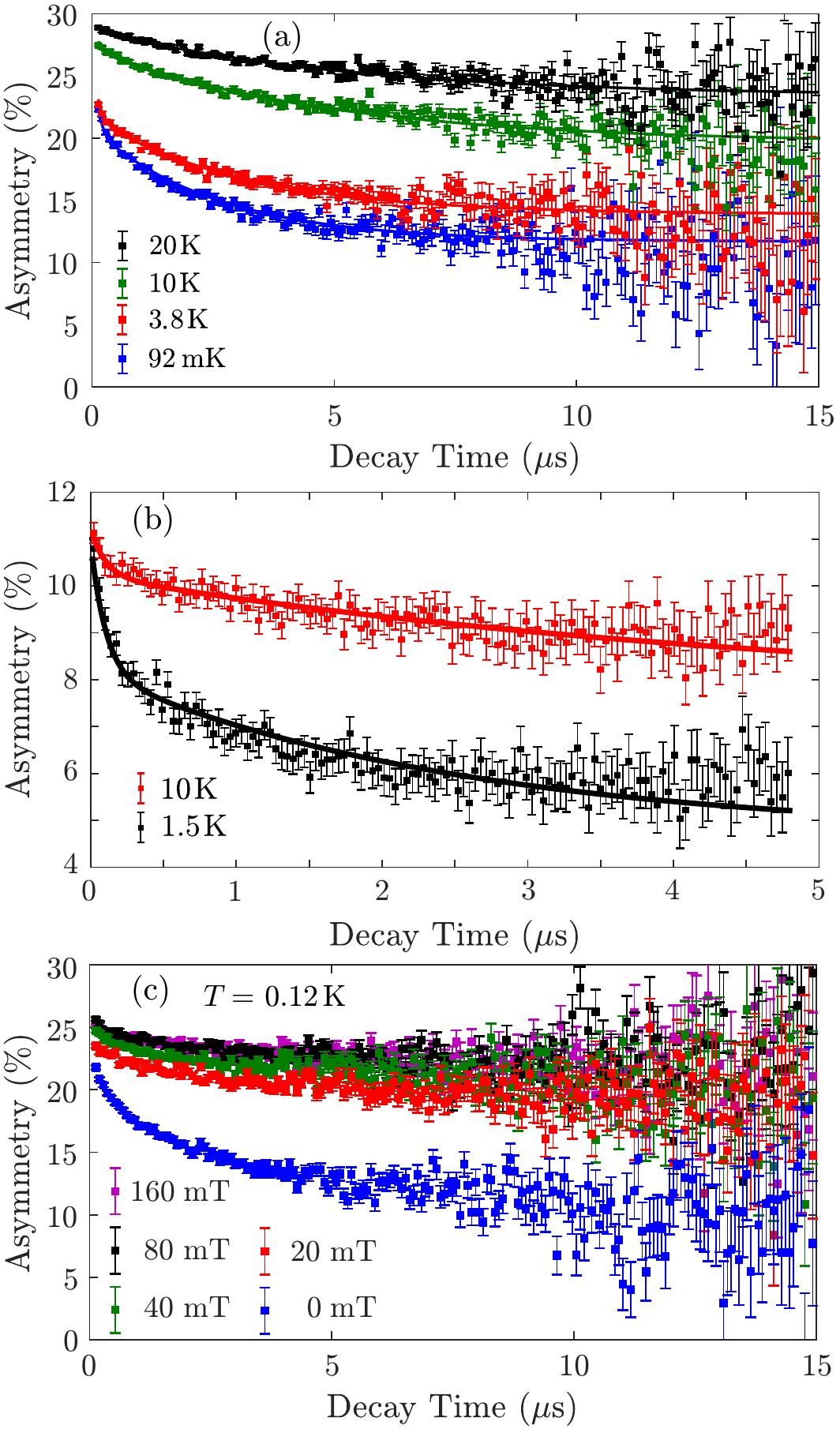}
	\caption{(a) Measured muon decay asymmetry as a function of decay time at selected temperatures for polycrystalline Y$_{2}$Os$_{2}$O$_{7}$. Datasets at temperatures greater than 3.8\,K were recorded in a $^{4}$He cryostat while those at 3.8\,K and below were taken in a dilution refrigerator. Solid lines represent a double-exponential fit as discussed in Section~\ref{MuSR_Analysis}. (b) Similar spectra to (a) but data taken at a different facility with higher time resolution to show that no structure has been missed in the low decay time region. This data was taken with the beamline's spin rotator switched on, leading to a lower absolute value of the measured asymmetry. (c) Muon decay asymmetry at 0.12\,K as a function of longitudinal applied field.}
	\label{Muon_Asym_Fig}
\end{figure}

\begin{figure*}
	\centering
	\includegraphics[width=0.98\textwidth]{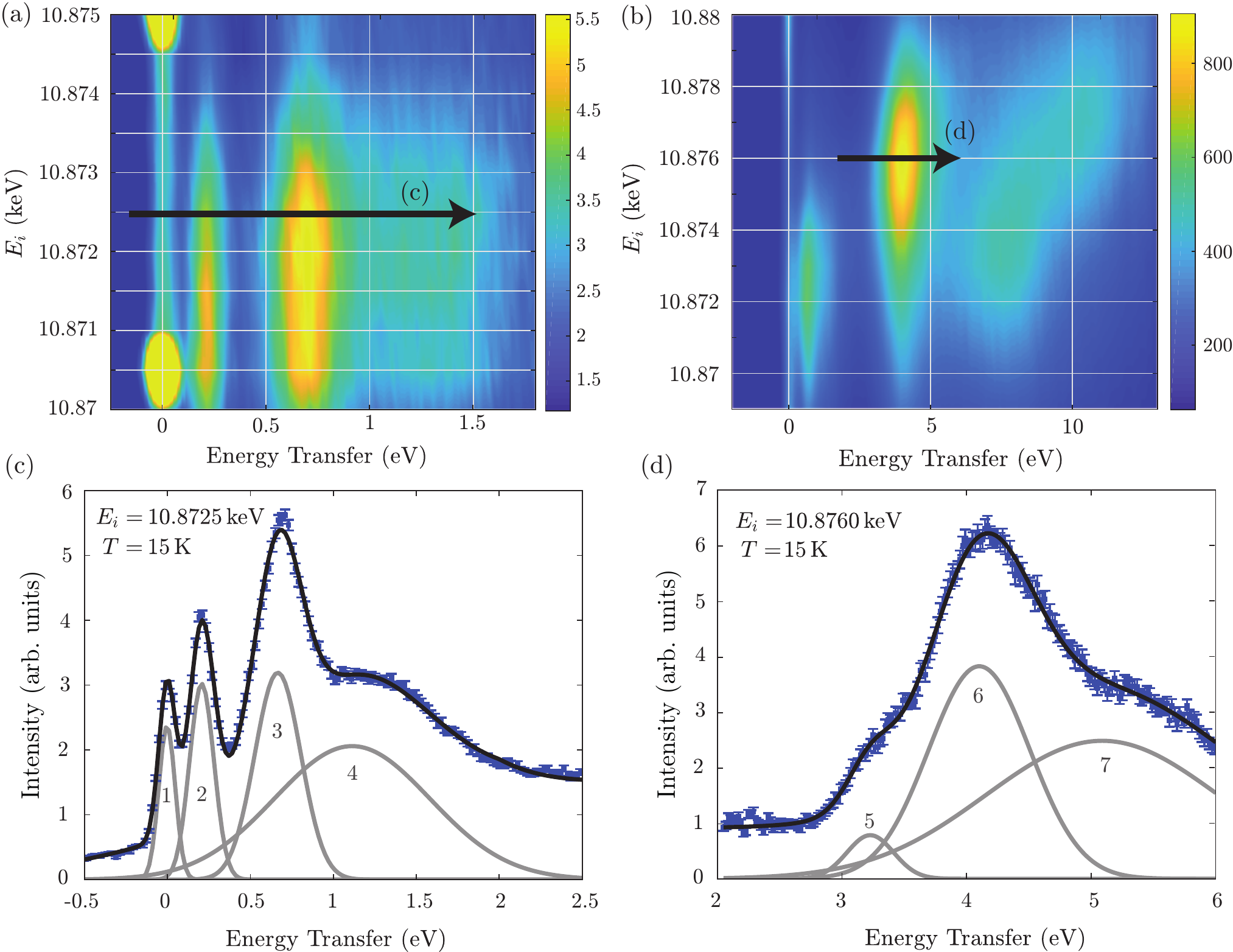}
	\caption{RIXS spectra of Y$_{2}$Os$_{2}$O$_{7}$ taken at 15\,K on the Os $L_{3}$ edge. (a) High resolution map focusing on the low energy excitations, (b) low resolution map, (c) a cut through the low energy excitations as marked in (a) and (d) a cut through the higher energy excitation as marked in (b). The black lines are a fit to four Gaussian peaks plus a linear background in (c) and three Gaussians plus a constant background in (d) with parameters shown in Table~\ref{Gaussian_Params_Table}. Gray lines show the fitted Gaussian components with numbers corresponding to peak numbers in the table. The very strong peaks at zero energy in the maps are Y$_{2}$Os$_{2}$O$_{7}$ Bragg peaks.}
	\label{RIXS_Data_Fig}
\end{figure*}

Zero-field $\mu$SR data measured at ISIS shows very little relaxation of the implanted muons at high temperatures $\gtrsim 100$\,K (not shown), as expected for a paramagnet, with relaxing behaviour developing gradually on cooling below this point [Fig.~\ref{Muon_Asym_Fig}(a)]. The relaxation becomes significantly greater below around 20\,K as the spin freezing temperature is approached, although the spectrum does not completely stop evolving even at the lowest measured temperature of 92\,mK. This indicates that the evolution of magnetic fluctuations in this system is very gradual, as is typical for spin glasses. No oscillations are seen at any temperature confirming that there is no long-range magnetic order, and additional datasets at 1.5\,K and 10\,K taken at PSI with much higher time resolution confirm that there is no oscillatory behaviour on shorter time scales down to 0.1\,$\mu$s [Fig.~\ref{Muon_Asym_Fig}(b)]. Overall the data are remarkably similar to those seen in canonical spin glasses such as AgMn \cite{Keren1996}, supporting the assertion that some kind of spin freezing occurs in this material.

In an applied longitudinal field at 0.12\,K [Fig.~\ref{Muon_Asym_Fig}(c)], a significant proportion of the relaxation is decoupled at the smallest measured field of 20\,mT, with no observable change between 80\,mT and 160\,mT. A similar longitudinal field dependence was also found at 2\,K (not shown). It has been shown that the relaxation caused by a distribution of static internal fields can be quenched by an applied field that exceeds the internal fields by about a factor of 10 (Ref.~\onlinecite{Hayano1979}). Therefore, our observations suggest that there is a small ($\sim 1$\,mT) static (on the muon precession timescale) component of the internal field in Y$_{2}$Os$_{2}$O$_{7}$.

\subsection{Resonant Inelastic X-Ray Scattering (RIXS)} \label{RIXSSection}

In Fig.~\ref{RIXS_Data_Fig}(a--b), we show RIXS maps of the Os $L_3$ resonance measured at 15\,K. In the lower resolution map (b) the most significant feature is a high intensity excitation peaked at energy transfer $\Delta E = 4.2$\,eV and $E_{i} = 10.876$\,keV. In the higher resolution map (a) two excitations are seen clearly at energy transfers $\Delta E = 0.2$\,eV and 0.7\,eV, as well as a weaker, broad feature around $\Delta E = 1.00$--1.25\,eV. Within the resolution of these data, all these lower energy excitations resonate at the same incident energy $E_i=10.8725$\,keV. There are also some broad, weak excitations at energy transfers around 3.33\,eV and above 5\,eV. Cuts through all these features [Fig.~\ref{RIXS_Data_Fig}(c--d)] show that no splitting into sub-levels is resolvable in any of them. 

As the incident photon energy is tuned to the Os $L_{3}$ edge we assume that the observed excitations involve Os $5d$ states. The crystal field at the Os site is close to cubic with a small perturbing trigonal distortion, so to a first approximation we can identify the $\Delta E = 4.2$\,eV feature with single-ion $t_{\rm 2g}$ -- $e_{\rm g}$ excitations and the low energy features with intra-$t_{\rm 2g}$ excitations. This assumption allows us to estimate the cubic crystal field parameter 10$Dq$ = 4.2\,eV. This assignment is supported by the fact that the $t_{\rm 2g}$ -- $e_{\rm g}$ and intra-$t_{\rm 2g}$ excitations resonate at energies separated by around 4\,eV and that this crystal field value is comparable to that found in other osmates, for example 10$Dq$ = 4.3\,eV in Ba$_{2}$YOsO$_{6}$ and 10$Dq$ = 4.5\,eV in Ca$_{3}$LiOsO$_{6}$ \cite{Taylor2017}. 

To quantify the energies and widths of these RIXS excitations, we performed phenomenological fits of the spectra in Fig.~\ref{RIXS_Data_Fig}, with the data in panels (c) and (d) modelled by a linear background and several Gaussian peaks. The corresponding fit parameters, numbered as indicated in the figures, are given in Table~\ref{Gaussian_Params_Table}.

\section{Analysis} \label{Analysis}

Our magnetization, heat capacity and $\mu$SR results provide evidence for spin glass behavior in Y$_2$Os$_2$O$_7$ with a small average magnetic moment per Os site of about 0.4 $\mu_{\rm B}$/Os, consistent with previous work (Ref.~\onlinecite{Zhao2016}). We shall now present analysis which shows that this moment is most likely associated with a small concentration of sites carrying large spins in a non-magnetic host, as opposed to a small spin on every site.

\begin{table}
	\centering
	\begin{tabular}{C{1.6cm}|C{1.6cm}|C{1.7cm}|C{1.6cm}}
		\hline
		\hline
		\\[-0.23cm]
		$i$ & $a_{i}$& $E_{i}$ (eV)& $\sigma_{i}$ (eV) \tabularnewline[3pt]
		\hline\\[-0.25cm]
		1 & 470(20) & -0.003(2) & 0.050(2) \tabularnewline[2pt]
		2 & 600(20) & 0.206(2) & 0.073(3) \tabularnewline[2pt]
		3 & 640(30) & 0.665(4) & 0.139(6) \tabularnewline[2pt]
		4 & 410(10) & 1.11(3) & 0.47(2) \tabularnewline[2pt]
		5 & 170(20) & 3.23(2) & 0.19(3) \tabularnewline[2pt]
		6 & 820(40) & 4.10(1) & 0.41(2) \tabularnewline[2pt]
		7 & 530(20) & 5.08(6) & 0.93(4) \tabularnewline[2pt]
		\hline
		\hline
	\end{tabular}
	\\[0.25cm]
	\caption{Gaussian parameters obtained from a fit of the sum four Gaussians plus a linear background to the RIXS spectra in Fig.~\ref{RIXS_Data_Fig}(c), $i = $ 1--4, and three Gaussians plus a constant background to Fig.~\ref{RIXS_Data_Fig}(d), $i = $ 5--7, where each Gaussian is of the form intensity~$=a_{i}\exp({-(E-E_{i})^{2}/2\sigma_{i}^{2}})$}
	\label{Gaussian_Params_Table}
\end{table}

\subsection{Field-Dependent Magnetisation}
	
\begin{figure*}
	\centering
	\includegraphics[width=0.98\textwidth]{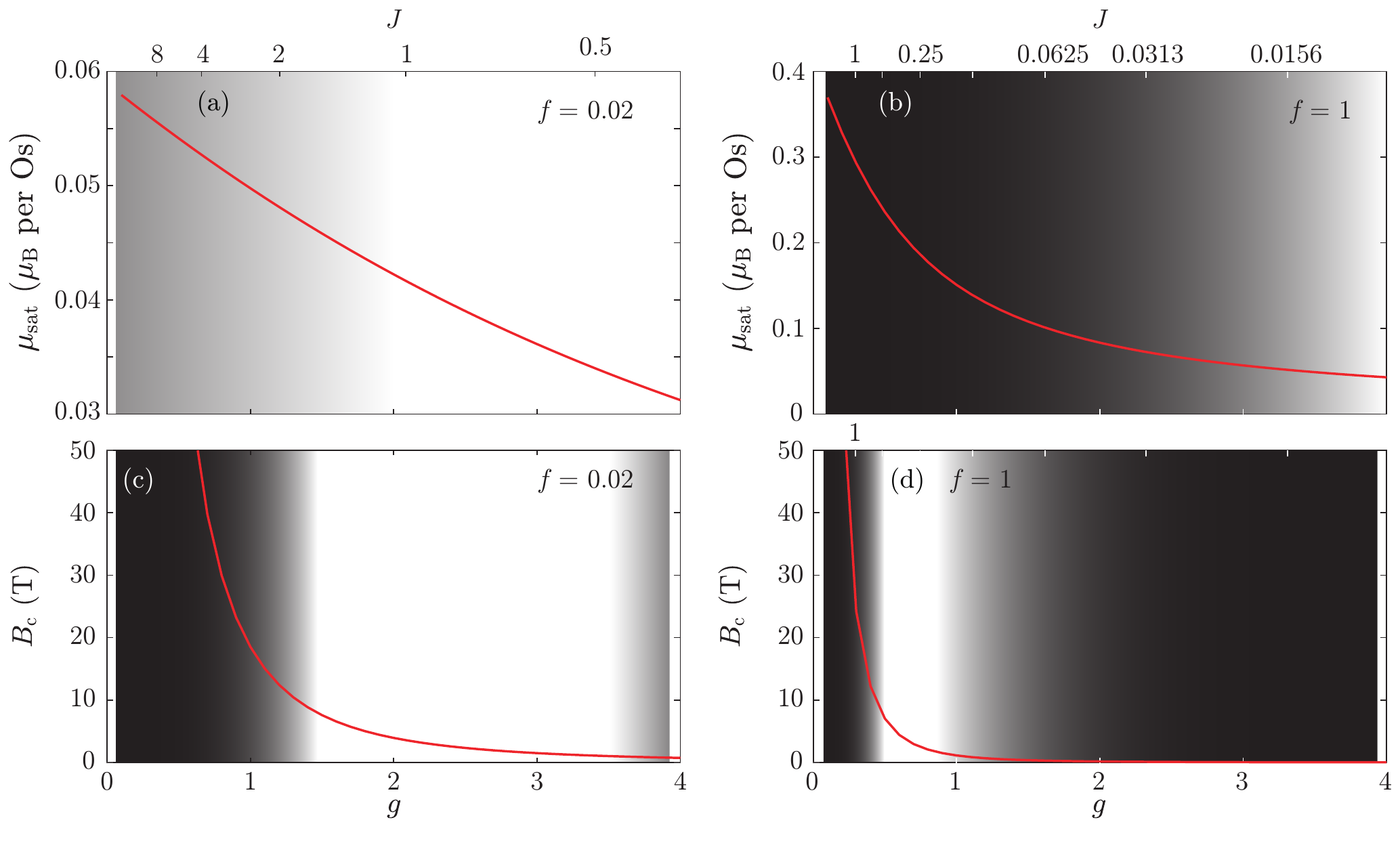}
	\caption{(a) and (b) the calculated saturated moment $\mu_{\rm sat} = fgJ\mu_{\rm B}$ as a function of $J$ (top axis) and $g$ (bottom axis) for two plausible values of $f$, found by solving Eqs.~(\ref{eq:2}) and (\ref{eq:3}) as described in the main text. (c) and (d) the calculated crossover field $B_{\rm c} = 5k_{\rm B}/gJ\mu_{\rm B}$ as a function of $g$ and $J$, found by solving Eqs.~(\ref{eq:2}) and (\ref{eq:4}) as described in the main text. The $g$ and $J$ axes are identical for plots at the same $f$. The grey shading indicates regimes of $g$ which imply saturation magnetizations (top panels) or saturation fields (bottom panels) that would be in poor agreement with our measurements (see text for a detailed discussion).}
	\label{MH_Analysis_Fig}
\end{figure*}
	
Assuming local moments with effective spin $J$, we can establish from the results of our Curie--Weiss fit that
\begin{equation}
	\sqrt{f} g \sqrt{J(J+1)} = 0.417(1),
	\label{eq:2}
\end{equation}
where $g$ is the $g$-factor of the moment. We are unable to determine $f$, $g$ and $J$ separately, but as the saturated moment 
\begin{equation}
	\mu_{\rm sat} = fgJ\mu_{\rm B}
	\label{eq:3}
\end{equation}
has a different dependence on $f$ and $J$ we can use the field dependence of our magnetisation data, together with assumptions derived from the observed spin-glass behavior, to test the likelihood of different values of $f$.
	
For this analysis we shall consider $f = 0.02$ and 1, chosen as representative of the scenarios in which the moments are dilute and concentrated, respectively.  The lower value $f = 0.02$ is typical of the levels of inter-site mixing and microscopic disorder reported in similar iridate materials \cite{Dey2016}.
	
For each $f$, we can use the constraint in Eq.~(\ref{eq:2}) to eliminate one of $g$ and $J$. Having done this, we can then calculate $\mu_{\rm sat}$ as a function of the remaining variable using Eq.~(\ref{eq:3}). In the discussion which follows, we shall assume $J$ has been eliminated in this way, leaving $\mu_{\rm sat}$ as a function of $g$ alone, however the analysis would proceed in the same way if we were to treat $\mu_{\rm sat}$ as a function of $J$ instead.

Plots of $\mu_{\rm sat}$ as a function of $g$ obtained this way are presented in Fig.~\ref{MH_Analysis_Fig}(a) for $f = 0.02$, and Fig.~\ref{MH_Analysis_Fig}(b) for $f = 1$. As discussed earlier, our magnetisation data indicate that $\mu_{\rm sat}$  does not exceed about $0.04$\,$\mu_{\rm B}$, and so from Figs.~\ref{MH_Analysis_Fig}(a--b) we see immediately that $g \gtrsim 2$ if $f = 0.02$, and $g \gtrsim 4$ for $f = 1$.
	
We now consider the implications of the spin-glass regime. Normal paramagnetic behavior is not observed here (see Fig.~\ref{MH_Fig}), and we assume that $\mu$ as a function of $H$ is governed instead by some average internal energy barrier $\Delta E$ comparable with the spin freezing temperature $T_{\rm f} = 5$\,K, i.e.~$\Delta E \simeq 5k_{\rm B}$.\footnote{We note that the authors of Ref.~\onlinecite{Zhao2016} have extracted an energy barrier to spin reorientation of $\Delta = 204(18)$\,K based on a fit of the Arrhenius Law $\omega = \omega_0\exp(-\Delta/T_{\rm f})$ to the peaks in the real part of the AC susceptibility. This fitting procedure has been shown to yield unphysically large energy barriers when applied to spin glasses, as discussed in Refs.~\onlinecite{Balanda2013, Huser1983, Huser1986, Souletie1985}, with information about $\chi''$ usually being required to obtain a physically realistic energy barrier.} At $T \ll T_{\rm f}$ we expect that spins can overcome the energy barrier and align with an external field $B$ provided $B \gg B_{\rm c}$, where 
\begin{equation}
	B_{\rm c}\simeq  \Delta E/(gJ\mu_{\rm B})
	\label{eq:4}
\end{equation}
is a crossover field. Figures~\ref{MH_Analysis_Fig}(c--d) plot $B_{\rm c}$ as a function of $g$ for $f=0.02$ and $f=1$, respectively, calculated from Eqs.~(\ref{eq:4}) and (\ref{eq:2}). We take $B_{\rm sat}/10$ as a lower limit for $B_{\rm c}$, where $B_{\rm sat} \simeq 10$\,T is an estimate of the saturation field at $T=0$ from Fig.~\ref{MH_Fig}, and we take the upper limit on $B_{\rm c}$ to be $B_{\rm sat}$. Hence, we conservatively estimate $B_{\rm c}$ to be in the range 1\,T $ \lesssim B_{\rm c} \lesssim $ 10\,T. The allowed range of $g$ corresponding to this acceptable range of $B_{\rm c}$ is indicated on Figs.~\ref{MH_Analysis_Fig}(c--d).
	
Figure.~\ref{MH_Analysis_Fig} shows that there are values of $g$ which are very improbable. These are represented by the darker shaded regions on the plots. For $f = 1$ we find that all values of $g$ are very unlikely under these constraints, while for $f = 0.02$ the constraints are satisfied when $1.5 \lesssim g \lesssim 4$. Converting this range of $g$ into a range of $J$ using Eq.(\ref{eq:2}), we obtain $0.5 \lesssim J \lesssim 1.5$. Overall we find that for $f = 0.02$ there is a wide range of physically reasonable $g$ and $J$ parameters consistent with the magnetisation data, but for $f = 1$ no such combination of parameters exists.
	
It is noteworthy that, since the calculated $\mu_{\rm sat}$ and $B_{\rm c}$ are both a factor of 5--10 larger for $f = 1$ than for $f = 0.02$, the above arguments still hold for quite significant changes in $f$ or in the experimental constraints. For example, any value of $f \lesssim 0.1$ still yields some plausible values of $g$, whereas any $f \gtrsim 0.5$ leads to all values of $g$ being unlikely based on experiment. We therefore conclude that the fraction of occupied Os sites is very likely to be on the order of a few \%, with the majority of Os sites adopting a non-magnetic state.

\subsection{Muon Spin Relaxation ($\mu$SR)} \label{MuSR_Analysis}

Having established that the spins in the sample are very likely to be dilute, we now perform fitting and simulations of the spin-glass-like relaxation in our $\mu$SR spectra.

\subsubsection{Fitting}

At all temperatures the $\mu$SR asymmetry appears to consist of a relaxing part plus a constant baseline component which does not relax even at long decay times, with the relative magnitudes of these two parts varying significantly with temperature. In order to quantify this we performed fits to the sum of two exponentials plus the baseline asymmetry, $A(\tau) = A_{\rm b} + A_{\rm r}((1-a)e^{-\lambda_{1}\tau}+ae^{-\lambda_{2}\tau})$ at each temperature where $A$ is the observed muon decay asymmetry, $\tau$ is the decay time, $\lambda_{1}$ and $\lambda_{2}$ are the two exponential decay rates, $A_{\rm r}$ is relaxing asymmetry due to muons experiencing a $B$ field in the sample, and $A_{\rm b}$ is the baseline asymmetry from muons which do not experience a magnetic field. $A_{\rm b}$ includes muons which stop in the sample holder, cryostat/dilution fridge and any non-magnetic or paramagnetic parts of the sample. Throughout this procedure the initial asymmetry $A_{\rm i} = A_{\rm r} + A_{\rm b}$ was held constant at $A_{\rm i} \simeq 30\%$ for the $^{4}$He cryostat and at $A_{\rm i} \simeq 28\%$ for the dilution fridge. These values were found by fitting the initial asymmetry to the highest temperature dataset available for each sample environment, since $A_{\rm i}$ should only be a function of the muon beam polarisation and sample environment and is not expected to change with temperature. For datasets above 40\,K we found that the spectrum fitted well to a single exponential so at these temperatures $a$ was fixed at 0, while for lower temperatures the full double-exponential form was required. The best possible fit was obtained by fixing $\lambda_2 = 9.66\,\mu{\rm s}^{-1}$ to its value in the lowest-temperature dataset in each sample environment for all double-exponential fits. 

Physically, exponential relaxation can result from a dilute distribution of static moments \cite{Walstedt1974} or dynamic moments with a single correlation time within the resolution of the spectrometer \cite{Khasanov2008}. The observation of two distinct exponential components may have a range of explanations including two different muon stopping sites, two different magnetic phases or two distinct correlation times for dynamic moments. Alternatively, the double-exponential fit may be a phenomenological fit to a more complex distribution of internal fields and relaxation times.

\begin{figure*}
	\centering
	\includegraphics[width=0.98\textwidth]{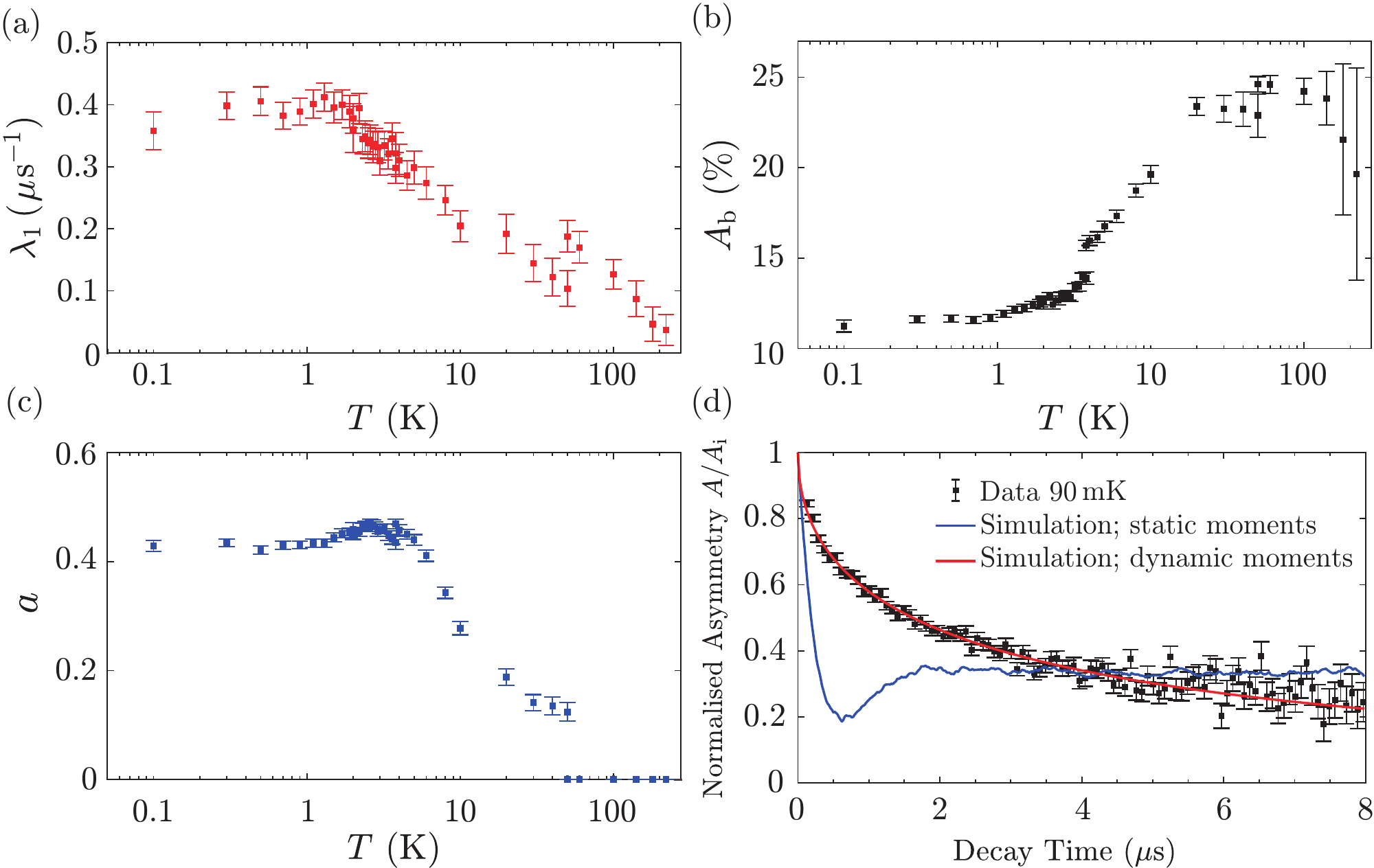}
	\caption{(a--c) Temperature dependence of the slower exponential relaxation rate $\lambda_1$, fraction of the faster relaxation $a$ and the background asymmetry $A_{\rm b}$ fitted to our $\mu$SR data by the procedure outlined in the main text. (d) Simulated $\mu$SR spectrum at 90\,mK for static and dynamic moments with moments of magnitude 2.95\,$\mu_{\rm B}$ on 2\% of Os sites.}
	\label{Muon_Sim_Fig}
\end{figure*}

The temperature dependences of the double-exponential fit parameters are presented in Fig.~\ref{Muon_Sim_Fig}. The relaxation rate $\lambda_{1}$ shows behaviour which is reminiscent of that seen in other spin glasses \cite{Keren1996, Uemura1985}, with an increase on cooling up to a peak at the spin-glass temperature followed by a plateau below this point. The spin freezing temperature $T_{\rm f}$ is $\simeq 3$\,K based on this measurement, which is slightly lower than that seen in AC susceptibility. This discrepancy may be due to the different fluctuation timescales probed by the different techniques.

The baseline asymmetry shows a clear decrease on cooling before flattening out below $T_{\rm f}$, except for a jump at 3.8\,K which can be attributed to the change of sample environment from $^{4}$He cryostat to dilution fridge at this temperature. This temperature dependence indicates that the volume of sample which is in a paramagnetic state decreases only gradually on cooling with no particularly sharp change at any temperature. A possible explanation for this would be if spins in different parts of the sample are freezing at slightly different temperatures, which is plausible behaviour for a spin glass.

Overall, it is very likely that the relaxing portion of the $\mu$SR spectrum is caused by the same part of the sample as the dilute spins which show hysteretic, spin-glass-like behaviour in magnetisation measurements. Since the muons can be assumed to stop randomly throughout the Y$_{2}$Os$_{2}$O$_{7}$ and impurity phases, we can therefore use the magnitude of the relaxing $\mu$SR signal $A_{\rm r}$ relative to the background $A_{\rm b}$ to examine which of the two phases the spins are located in.

Muons stop randomly in the impurity and Y$_{2}$Os$_{2}$O$_{7}$ phases in proportion to their volume, and it is reasonable to assume that muons stopping in the Y$_{2}$Os$_{2}$O$_{7}$ would not couple significantly to magnetic fluctuations in the impurity phase. Since the impurity is only a few \% of the sample by volume, the relaxing asymmetry $A_{\rm r}$ would therefore be much smaller than the baseline asymmetry $A_{\rm b}$ if the impurity were the source of spin-glass behaviour. At low temperature, the baseline asymmetry $A_{\rm b}$ is similar in value to the relaxing asymmetry $A_{\rm r}$ (e.g. $A_{\rm r} = 13.4$\,\% and $A_{\rm b} = 11.4$\,\% at 92\,mK). We therefore find that a magnetic impurity phase cannot be the source of the observed spin-glass behaviour.

If the spins are located in the Y$_{2}$Os$_{2}$O$_{7}$ phase, magnetic exchange mediated by ions located between the spins and the muon stopping sites in this phase would likely cause a significant magnetic field at the muon site. For example, in the double perovskite iridates it has been shown that exchange mediated by Y$^{3+}$ and O$^{2-}$ ions is significant even between second- and third- nearest-neighbour Ir sites \cite{Fuchs2018}. Furthermore, simulations presented in Ref.~\onlinecite{Fuchs2018} show that for a double perovskite lattice populated with a few \% spins on one of the octahedral sites the majority of Ir sites are no further than the third-nearest-neighbour distance from a spin. Assuming that similar results hold for the pyrochlore structure of Y$_2$Os$_2$O$_7$, muons stopping at most locations within the Y$_{2}$Os$_{2}$O$_{7}$ phase will experience a significant magnetic field even if the spin concentration is low. We therefore conclude that the relaxing behaviour is consistent with dilute moments in the main Y$_{2}$Os$_{2}$O$_{7}$ phase, and that the impurity phase shows no noticeable signal other than a constant background in $\mu$SR.

\subsubsection{Simulations}

For the lowest temperature dataset we have performed a simulation similar to that presented in Ref.~\onlinecite{Kirschner2016} to try to extract information about the spin dynamics. This simulation involves randomly populating a lattice with magnetic moments $\mu$ on a fraction $f$ of the sites then examining the internal field at a muon test site.

If the spins are assumed to be completely static, the simulation results in an asymmetry 
\begin{equation}
A(t) = \int p(\Delta)(\frac{1}{3} + \frac{2}{3}\cos{(\gamma_{\mu}\Delta \tau)}){\rm d}\Delta
\end{equation}
where $\Delta/\gamma_{\mu}$ is the width of the field distribution at the muon site, $p(\Delta)$ is the probability of finding that field width for a randomly chosen muon site and $\gamma_{\mu} = 2\pi \times 135.5$\,MHz\,T$^{-1}$ is the gyromagnetic ratio of the muon \cite{Hayano1979}. The simulated spectrum assuming the most likely values of $f = 0.02$ and $\mu = 2.95$\,$\mu_{\rm B}$ is presented in Fig.~\ref{Muon_Sim_Fig}(d), however we find that this model cannot reproduce the data for any values ($f,  \mu$).

If the spins are allowed to fluctuate, the model asymmetry becomes 
\begin{equation}
A(t) = \int p(\Delta){\rm e}^{-2\Delta^2 \tau/\nu}{\rm d}\Delta
\end{equation}
where $\nu$ is the fluctuation rate \cite{Hayano1979}. The model with fluctuations provides a much better fit to the data. For $\mu \simeq 2.95\,\mu_{\rm B}$ and $f \simeq 0.02$ the best fit is achieved with $\nu \simeq 21$\,MHz, as plotted in Fig.~\ref{Muon_Sim_Fig}(d). This fit is, however, reliant on an adjustment of $A_{\rm b}$ from the previously fitted value of 11\% to $\simeq 7$\%. If the baseline asymmetry is fixed at 11\% we find that the model cannot reproduce the data even in the dynamical case. This indicates either that the data is not well-modelled by this scenario or that even at the lowest temperatures a significant fraction ($\simeq 18$\%) of muons stopping in the sample experience a non-magnetic or paramagnetic environment. Given that we have ruled out impurities on the $>10\%$ level this latter situation would imply that the Y$_2$Os$_2$O$_7$ phase still contains some non-magnetic regions even well below $T_{\rm f}$.

\subsection{RIXS Single Ion Calculations} \label{RIXS_Single_Ion}

\begin{figure*}
	\centering
	\includegraphics[width=0.8\textwidth]{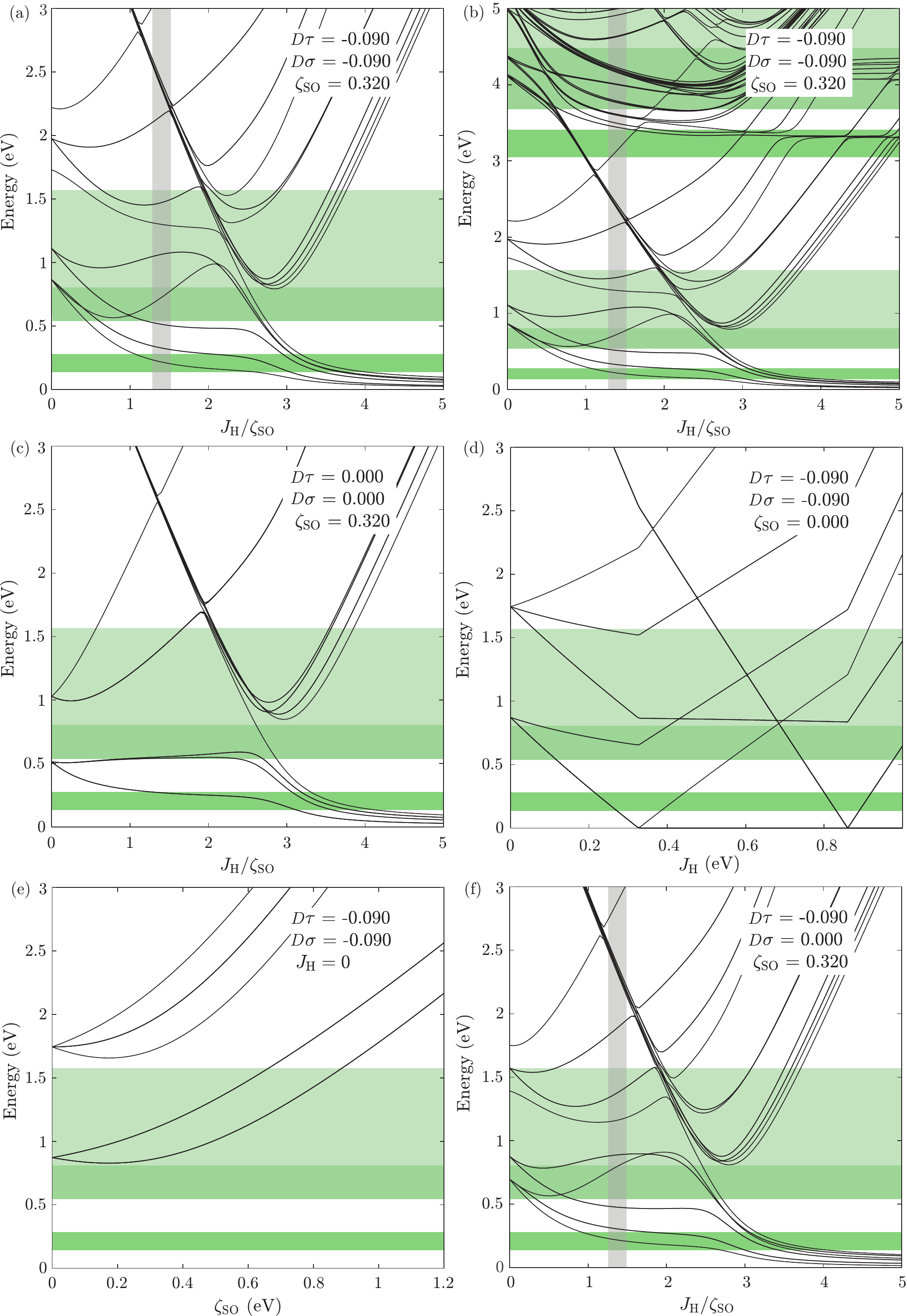}
	\caption{The predicted energy levels of Os 5$d^{4}$ electrons from single-ion calculations as outlined in the main text. Green horizontal strips represent the observed energy levels in the present RIXS experiments with the position and width of the strip corresponding to the Gaussian peak position and full width at half maximum from Table~\ref{Gaussian_Params_Table}. Vertical grey strips in (a) and (b) indicate a region where the calculated low energy levels appear to best match those seen in experiment as discussed in the main text, while the grey strip in (f) is in the same position as those in (a--b) to aid visual comparison. (a) Calculation with non-zero trigonal distortion, spin-orbit interaction and Hund's coupling. (b) Same as (a) but extended to high energy transfer. (c) Calculation with no trigonal distortion. (d) Calculation with no Hund's coupling. (e) Calculation with no spin-orbit interaction. (f) Calculation with the variable trigonal distortion parameter $D \sigma$ set to zero.}
	\label{RIXS_Fitting_Fig}
\end{figure*}

It is reasonable to assume RIXS is sensitive to all Os sites in the sample, the majority of which we have now established to be non-magnetic based on the above analysis of our other experimental results. We therefore performed single-ion calculations assuming the $d^4$ electronic configuration for the Os ions, including inter-electron interactions, spin-orbit interaction and trigonal crystal field terms in the Hamiltonian in order to understand the origin of the excitations seen in RIXS. This procedure is outlined in Refs.~\onlinecite{Gerloch1983, Konig1977} and involves writing each contribution to the Hamiltonian as a matrix using the properly antisymmetrized multielectron states of the $d^{4}$ configuration as a basis then numerically diagonalising the combined Hamiltonian. The inter-electron interaction is written in terms of Racah parameters \cite{Georges2013} $A$, $B$ and $C$ which can be transformed into intra- and inter-orbital Coulomb interactions $U$ and $U'$ and the effective Hund's coupling $J_{\rm H}$ via
\begin{equation}
\begin{split}
J_{\rm H} &= 3B + C \\
U &= A +4B +3C \\
U' &= A - 2B + C.
\end{split}
\end{equation}
Following Ref.~\onlinecite{Hempel1976} the crystal field is parametrised by $Dq$, $D \sigma$ and $D \tau$, where $Dq$ represents the octahedral crystal field, and $D \tau$ and $D \sigma$ small trigonal distortions away from the perfect octahedral case.~\footnote{The parameters $Dq$, $D \sigma$ and $D \tau$ used in this work correspond to the parameters with the same symbols in Ref.~\onlinecite{Hempel1976}.} The spin-orbit coupling strength enters via a single parameter $\zeta_{\rm SO}$.

Some of the above parameters could be found from experiment before performing calculations. We have estimated from the RIXS data that 10$Dq$ = 4.2\,eV, and there is a direct relationship between $D \tau$, the sign of $D \sigma$ and the $48f$ oxygen position $x = $ 0.3352(2) \cite{Lever1972, Hempel1976b} which yields $D \tau = -0.090$\,eV and tells us that $D \sigma$ must have the same sign (--) as $D \tau$. The Racah parameter $A$ only appears on the diagonal elements of the Hamiltonian and causes only a constant shift of all energy levels. Since spectroscopy reveals only relative, not absolute energies, $A$ is not determined by this measurement.

All other parameters ($B, C, \zeta_{\rm SO}$ and $|D \sigma|$) are in general free and ideally would be fitted to experimental data. Unfortunately, we do not observe enough excitations in the experiment for this to be possible in this case. Instead, we need to fix some of the parameters to values obtained from other, similar compounds.

We fixed the values of $\zeta_{\rm SO} =$ 0.32\,eV and $B = 0$ to the values obtained for Ba$_{2}$YOsO$_{6}$ in Ref.~\onlinecite{Taylor2017} leaving as free parameters $D \sigma$ and $C = J_{\rm H}$. Fig.~\ref{RIXS_Fitting_Fig}(a--b) shows the results of this calculation, in which best agreement with experiment is obtained for $D \sigma = -0.09$\,eV and a region around $J_{\rm H}/\zeta_{\rm SO} \simeq 1.4$. All of the experimental peaks are consistent with one or more excitations in the calculation, although notably many of the experimental features are in fact a combination of several closely-spaced levels which are unresolved. There are several states (for example those at 0.3\,eV and 0.5\,eV) which fall close to the edges of the experimental peaks, as well as one set of nearly-degenerate states at around 2.1\,eV which is not close to any feature in the experiment. It is possible that these states may have a low spectral weight if a full RIXS calculation were performed, in which case they would be unresolvable above the background, especially in the case of the 0.3\,eV and 0.5\,eV levels which may easily be swamped by the nearby, stronger excitations or combined with them via intersite hopping terms which are not included in this model.

Changing $D \sigma$ causes small perturbations to the low-lying energy levels and makes the agreement with experiment less good. The $D \sigma = 0$ case is presented in Fig.~\ref{RIXS_Fitting_Fig}(f) for comparison, showing how agreement is still close but slightly worse, in particular for the two lowest-energy observed peaks.

The above results are consistent with work on other osmates and iridates which has found $J_{\rm H} \simeq \zeta_{\rm SO}$ \cite{Calder2016, Taylor2017, Yuan2017}. We emphasise that because of the number of free parameters and the inherent uncertainty due to the unknown RIXS matrix elements we cannot conclude that the parameter values suggested here are definitely the values in this material, only that based on our current knowledge the present model is capable of explaining the data for plausible values of all parameters.

We also performed calculations by removing in turn the trigonal distortion ($D \tau = D \sigma = 0$, Fig.~\ref{RIXS_Fitting_Fig}(c)), spin orbit interaction ($\zeta_{\rm SO} = 0$, Fig.~\ref{RIXS_Fitting_Fig}(d)) and inter-electron interactions ($B = C = 0$, Fig.~\ref{RIXS_Fitting_Fig}(e)). In all three cases we could not find any values of the remaining parameters which adequately reproduce the two lowest lying features at 200\,meV and 700\,meV in the RIXS spectrum. This allows us to conclude that all three effects (trigonal distortion, spin--orbit coupling and inter-electron interactions) are required to model the physics of this material.

The ground state for our likely set of parameters, as well as for \textit{any} set of parameters calculated here as long as $J_{\rm H}/\zeta_{\rm SO} \lesssim 3$, is a $J_{\rm eff} = 0$ non-magnetic singlet. This allows us to rule out any kind of single-ion physics, including the trigonal distortion, as the source of the magnetic moment, consistent with our conclusions from other techniques. 

For all reasonable sets of parameters there are low-lying excitations in the 200--400\,meV region, which may be either a doublet, triplet or closely-spaced singlet and doublet. Significantly, we find that for the trigonally distorted parameters found here ($D\tau = D\sigma = -0.09$\,eV) these lowest-lying excitations are a singlet at 200\,meV and a degenerate doublet at 300\,meV. This is in contrast to the undistorted case where the first excitation is a triplet. Quantitative theories of excitonic magnetism applied to the $A_{2}$YIrO$_{6}$ ($A = $Sr, Ba) materials such as Ref.~\onlinecite{Khaliullin2013} are based on a situation where the first excited state is a low-lying triplet. In the scenario proposed here for Y$_2$Os$_2$O$_7$ the splitting between the singlet and doublet excitations is $\simeq 50\%$ of the separation between the singlet excitation and the ground state, representing quite a significant departure from the scheme used in the theories. The theories may therefore need modification before being directly applied to the pyrochlore osmates.

In light of these calculations, we can now also explain the temperature independent component of the magnetic susceptibility in Fig.~\ref{DC_Fig}. It was shown in Refs.~\onlinecite{Chen2017, Khaliullin2013} for Ba$_{2}$YIrO$_{6}$ that, following standard second order perturbation theory, in the single-ion case the magnetic susceptibility of a system with a singlet ground state and a low-lying triplet excited state at 350\,meV is temperature-independent and on the order of $\chi_{0} \sim 1\times10^{-3}$\,emu mol$^{-1}$. For Y$_2$Os$_2$O$_7$, the first and second excited states are a singlet and a doublet, respectively, with a similar average energy above the ground state as the triplet in Ba$_{2}$YIrO$_{6}$. The van-Vleck susceptibility for Y$_2$Os$_2$O$_7$ is therefore expected to be of similar magnitude to that of Ba$_{2}$YIrO$_{6}$, consistent with our observed value of $8.96(2)\times10^{-4}$\,emu\,mol$^{-1}$.

\section{Discussion}
\label{Discussion}

When all of our experimental results are considered together, a consistent picture emerges with the majority of Os sites in a non-magnetic $J_{\rm eff} = 0$ state along with a few sites exhibiting a large spin. These magnetic defect sites are likely caused by some kind of microscopic disorder, for example related to oxygen deficiency in the sample, site disorder involving partial interchange of Y$^{3+}$ and Os$^{4+}$ ions, or partial static charge disproportionation (2\,Os$^{4+} \to $ Os$^{3+} + $ Os$^{5+}$).

This scenario is very similar to that proposed in the recent preprint Ref.~\onlinecite{Fuchs2018} for the $5d^{4}$ iridate Ba$_{2}$YIrO$_{6}$, where the authors show via electron spin resonance (ESR) spectroscopy that the observed magnetic moment is caused by a small percentage of Ir$^{6+}$ ($5d^{3}$) and Ir$^{4+}$ ($5d^{5}$) magnetic defects, with the majority of Ir sites remaining in the non-magnetic Ir$^{5+}$ ($5d^{4}$) configuration. A similar scenario in Y$_{2}$Os$_{2}$O$_{7}$ would be consistent with all of our data; for example, only 1\% of Os sites in the spin-only $5d^{3}$ configuration ($L = 0$, $J = S = 3/2$) would lead to $\sqrt{f}\mu_{\rm eff} = 0.387\,\mu_{\rm B}$, very close to our measured value of $\sqrt{f}\mu_{\rm eff} = 0.417\,\mu_{\rm B}$.

The authors of Ref.~\onlinecite{Fuchs2018} also show that medium- and long-range interactions, possibly involving exchange mediated by Y ions, are significant in Ba$_{2}$YIrO$_{6}$ and that the magnetic defects tend to form extended correlated clusters even at low concentrations. Such long-range interactions and clustering of magnetic defects would provide a natural explanation for the spin-freezing in Y$_{2}$Os$_{2}$O$_{7}$, including the observation in our $\mu$SR that the proportion of the sample exhibiting non-magnetic behaviour reduces gradually with temperature, and that some regions of the sample appear non-magnetic even below $T_{\rm f}$.

\section{Conclusion}

Our AC and DC magnetisation, heat capacity and $\mu$SR measurements all show results consistent with low temperature spin-glass behaviour as suggested in Ref.~\onlinecite{Zhao2016}. Having ruled out impurity effects, we have shown that the observed Curie-Weiss-like moment in Y$_{2}$Os$_{2}$O$_{7}$ is very likely due to large moments $\simeq 3\,\mu_{\rm B}$ located on a small proportion $f \simeq 0.02$ of Os sites, perhaps related to microscopic disorder in the sample. We have also shown via RIXS measurements in conjunction with single-ion energy level calculations that the majority of Os sites in Y$_{2}$Os$_{2}$O$_{7}$ exhibit a $J_{\rm eff} = 0$ ground state with low-lying singlet and doublet excited states in the single-ion picture. Overall, a scenario similar to that recently proposed by Fuchs {\it et al.} \cite{Fuchs2018} with a small proportion of magnetic defects in a $5d^{3}$ or $5d^{5}$ configuration along with majority non-magnetic $5d^{4}$ Os sites can explain all of our observations.

\section{Acknowledgements}

We thank F. Lang for useful discussions. The European Synchrotron Radiation Facility (ESRF) provided synchrotron radiation. Experiments at the ISIS Neutron and Muon Source were supported by a beamtime allocation from the Science and Technology Facilities Council. This work was supported by the U.K. Engineering and Physical Sciences Research Council (Grant Nos. EP/N034872/1 and EP/N034694/1, and a studentship for N.R.D.), the National Natural Science Foundation of China (11774399，11474330), the Chinese Academy of Sciences (XDB07020100 and QYZDB-SSW-SLH043) and the Shanghai Pujiang Program (17PJ1406200). F.~K.~K.~K. thanks Lincoln College, Oxford, for a doctoral studentship.

\bibliographystyle{apsrev4-1}
\bibliography{Davies_Y2Os2O7}

\end{document}